\documentclass[showkeys,showpacs,pre,preprint,final]{revtex4}

\begin{document}

\title{Resolving Paradoxes of Classical Nucleation Theory} 
\author{Yossi Farjoun}
\author{John Neu}
\affiliation{University of California, Berkeley}
\date{\today}

\begin{abstract}
We present a new model of homogeneous aggregation that contains the
  essential physical ideas of the classical 
  predecessors, the Becker-D\"oring and Lifshitz-Slyovoz models.
  These classical models, which give different predictions, are
  asymptotic limits of the new model at small (BD) and large (LS)
  cluster sizes.  
  Since the new theory is valid for large and small clusters, it allows
  for a complete description of the nucleation  process; one that can
  predict the creation of super-critical clusters at the Zeldovich
  nucleation rate, and the diffusion limited  growth of large clusters
  during coarsening. 
  By retaining the physically valid ingredients from both models, we
  explain the seeming incompatibilities and arbitrary choices of the
  classical models. 
\end{abstract}



\maketitle
\section{Introduction}
Nucleation refers to the aggregation of identical particles (monomers) into clusters. 
Its universality throughout physics, chemistry and biology is well known. 
References \cite{KGT83}, \cite{KELT91}, \cite{NB00}, \cite{GWS01},
\cite{ISRAELACHVILI91}, \cite{NCB02}, \cite{LS61}, \cite{XH91},
\cite{MG96}, \cite{GNON03} provide a lineup of the `usual' (and some unusual) suspects.
Also well known are the long-standing challenges that aggregation poses to modeling.
Two classical models of aggregation due to Becker-D\"oring (BD) \cite{BD35}, and Lifshitz-Slyozov (LS) \cite{LS61} are incomplete and mutually inconsistent.

In BD, clusters exchange particles with the surrounding monomer bath by a `surface reaction', and it is assumed that the monomer bath around the clusters has uniform concentration.
This is only possible with infinite diffusivity of monomers. 
While this description is asymptotically accurate for sufficiently small clusters,
 the uptake of monomers by large clusters is strongly controlled by  the diffusivity.
LS describes cluster growth and shrinkage controlled by  diffusion of monomers.
In LS, the monomer concentration at the surface of a cluster is a prescribed function of the local curvature, generally different from the `background' concentration, far from clusters.
Hence, monomer concentration about a large cluster is nonuniform, and there is diffusive transport of monomer into or away from the cluster.
This physics of LS leads to a prediction for cluster growth that disagrees with BD\@.
Furthermore, LS is `incomplete', in that it does not describe the initial creation of clusters from pure monomer.
While it is generally accepted that BD is a model for small clusters, and
  LS  for large, several questions remain.
How to interpolate between the two models?
What is the characteristic size that separates `large' from `small'? 
What physics governs the growth in the intermediate scale?
What globally valid model encompasses the whole evolution of clusters,
  from an initial state of pure monomer to the asymptotic self-similar distribution of large cluster sizes?

The current paper  presents a new model that retains the essential
physical ingredients: the clusters gain and lose monomers by a surface reaction  that depends on the cluster size and the monomer concentration \emph{seen on the surface}.
Monomers outside the cluster undergo diffusion with finite diffusivity.
These ingredients give rise to a free boundary problem for the
growth of a cluster that contains  a new intrinsic cluster size,
$k_*$, in addition to the well known \emph{critical size}, $k_c$.
The critical size, $k_c$, separates shrinking clusters ($k<k_c$) from growing ones ($k>k_c$).
The new cluster size, $k_*$, indicates the importance of diffusion: 
the new prediction for cluster growth asymptotes to BD for small clusters with $k\ll k_*$, and to LS for large clusters with $k\gg k_*$.
In the former case, the diffusion effectively equates the surface density of monomer with the far-field density, thus, the surface reaction dictates the growth.
In the latter case, growth is strongly limited by finite diffusivity.
Furthermore, the  new model of cluster growth interpolates between BD and LS for intermediate cluster sizes on the order of $k_*$.

The smooth interpolation between BD and LS is crucial for a global model of aggregation that describes the whole process, from the initial creation of clusters from pure monomer, to the late stage growth-attrition process called \emph{coarsening}.
The essential idea is simple: if $k_c\ll k_*$, as expected in most cases, standard BD describes the \emph{nucleation} of super-critical ($k>k_c$) clusters and their growth while $k_c<k\ll k_*$. 
The super-critical clusters rapidly grow to sizes $k\gg k_*$, and their subsequent careers are described by LS.

Mathematically, we model this physics by a continuum approximation of
the discrete kinetics. 
The continuum equations constitute a PDE signaling problem for the distribution $r(k,t)$ of large ($k\gg k_*$) clusters in the space of (continuous) cluster size $k$.
At the lowest order of approximation, the cluster-size distribution satisfies an advection PDE, in which the growth rate ($\dot k$ vs. $k$) furnishes  the advection velocity.
The classical Zeldovich formula \cite{ZELD43}, which follows from BD, computes the creation rate  of super-critical ($k>k_c$) clusters.
Since we assume $k_c\ll k_*$, the Zeldovich formula gives rise to an effective source boundary condition on $k=0$.
The initial state of  pure monomers is expressed by a  zero initial condition, $r(k,0)\equiv 0$. 
Information about the amount of small ($k<k_c$) clusters is not expressed directly in $r(k,t)$. 
Instead, using conservation of particles, we express the amount of sub-critical clusters using an integral of $r(k,t)$.

Our theory does not handle nucleation that happens with $k_c$ on the order of $k_*$.
For this we suspect that a new theory is needed, one that considers the discrete and fluctuating nature of the monomer bath, and does not resort to
the diffusion equation, which arises from mean-field averaging.

The paper is structured as follows.
In section \ref{sec:BD} we present a short summary of the classical microscopic
aggregation theory (BD). 
We derive rate constants for attachment and dissociation of monomers from a cluster by using free energy and detailed balance arguments.
The only difference from the classical theory is that it is based on
the \emph{surface} density of monomers, the density of monomers just outside the cluster, and does not assume that the monomer density is homogeneous.

In section 3 we take into account the finite spatial diffusion of
monomers. 
While still focusing on a single
cluster, we connect the surface monomer density with a far-field monomer
density, the nearly uniform concentration of monomers far from any cluster. 
This prescribes the growth rate of a cluster as a function of its size and the far-field monomer density. 
The standard assumption in diffusion limited aggregation is that the
surface density corresponds to a critical  cluster i.e.,
growing and shrinking are equally likely.
This seems paradoxical for two reasons. 
First, the free energy of a cluster as a function of cluster size, $k$, has its global maximum at $k=k_c$. 
On the face of it, this seems to be an unstable equilibrium, but yet it is claimed that the cluster remains at the top of this equilibrium.
In addition, if that is the value of the monomer density, how
does the cluster grow or shrink? 

This paradox is another artifact of assuming uniform monomer concentration as in BD\@.
It is deconstructed at the end of section 3, by an asymptotic analysis which exposes the stabilizing role of finite monomer diffusion:
If monomer concentration at the surface of a large ($k\gg k_*$) cluster has large deviation from the critical value described above, the surface reactions rapidly absorb or expel monomers.
Consequently, due to finite diffusivity, the surface monomer concentration undergoes a collateral adjustment \emph{towards} the critical value and the rapid reactions are turned off.
The growth rate due to diffusion is much slower and dictates the evolution of large clusters.

In section \ref{sec:zeldovich} we turn to the \emph{ensemble} of all
clusters.
If the density of monomers is below a certain \emph{saturation
  value},  an equilibrium exists, in which large clusters are extremely
unlikely.
For a `super-saturated' ensemble, with monomer density exceeding the
saturation value, clusters greater than a `critical cluster size',
$k_c$, have a strong tendency to persist and grow.
In this super-saturated case, there is no equilibrium; the
distribution of clusters is continuously changing.
Initially, there is \emph{nucleation}, which is the creation of
super-critical clusters. 
The calculation of the nucleation rate based on BD is reviewed here. 

As stated before, this paper proposes a PDE signaling problem for the distribution of cluster sizes that quantifies the complete evolution of the aggregation process.
Section 5 contains the assembly of the signaling problem from the component
parts in sections 2--4.
It has a peculiar nonlinearity, in which the advection velocity in the PDE and the boundary condition (BC) at $k=0$ depend on the monomer
density as a parameter.
The monomer density can be written as an integral of the solution, and herein lies the nonlinearity.

The nonlinearity makes the task of solving the equations difficult
enough to warrant placing it in a separate, forthcoming  paper.
\section{Classical Becker-D\"oring Model}
\label{sec:BD}
Becker-D\"oring theory (BD) imposes simplifying assumptions at the outset:
The clusters are assumed to be uniformly distributed in a dilute
`bath' of monomers. 
This assumption is adjusted regarding the distribution of
monomers in the next section, when we add diffusion. 
The clusters are assumed to change size only by losing or gaining
one monomer at a time. 
Two large clusters do not fuse together nor does one cluster
 break into two. 
This can be justified heuristically by noticing that the density of
the large clusters is much smaller than the (already small) density of
monomers, thus the probability of two large clusters interacting is
small. 
In addition, the mobility of the large clusters is much smaller than that of the monomers, so they are even less likely to stumble upon one another. 
Similarly, since the large clusters have a low mobility (relative to monomers) a cluster that breaks into two will, most likely, reconnect quickly, as the two parts remain close together. 

Another important assumption is that the only governing parameter of a cluster is its size. 
The shape of the cluster is assumed to be fixed. 
This assumption can be weakened to require that clusters of same size have the same binding energy and the same surface area.

To derive the kinetic model of nucleation we introduce the essential quantitative ingredients: energy,
free energy, and the rate constants of transitions between configurations.
\subsection{Energy and Free Energy}

\begin{figure}[ht!]
\centerline{
\resizebox{6cm}{!}{\includegraphics{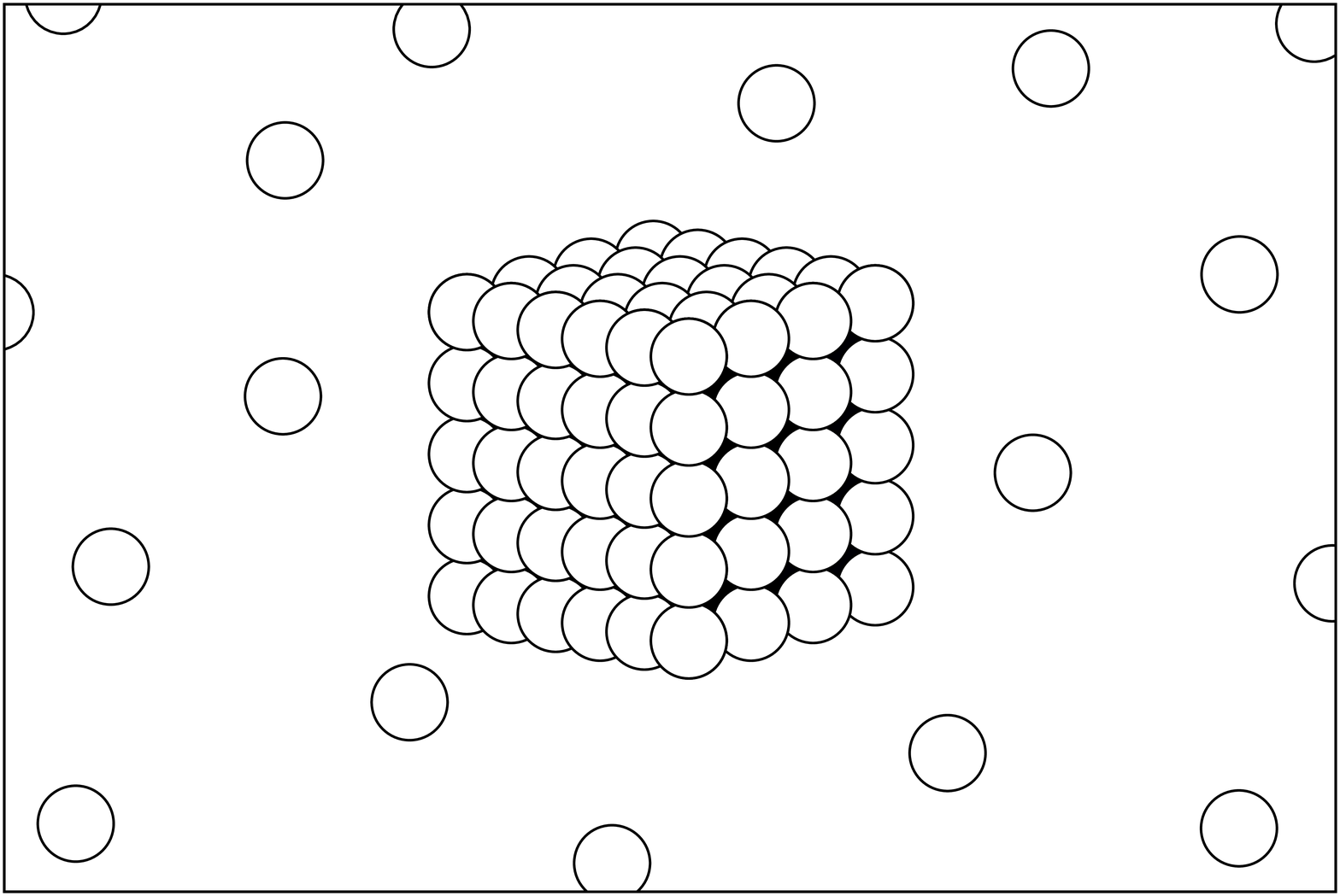}}}
\caption[Schematic cubic crystal]{A schematic cubic crystal surrounded by the monomer bath.}
\end{figure}
The energy of cubic cluster with $k$ monomers is
\begin{equation}
  \eps_k \sim 3 \eps (k-k^\frac23).\label{eq:energy:cube}
\end{equation}
Here $\eps$ is the binding energy of a single bond between two
adjacent particles, the energy needed to break it. 

While the clusters are not assumed to be simple cubes,  the general structure of the binding energy is expected to remain. 
Thus, we assume that there is a bulk energy constant, $\alpha>0$, and a surface energy constant, $\sig>0$, such that the energy of a $k$-cluster is
\begin{equation}
  \eps_k \sim k_BT(\alpha k-\frac32\sig k^\frac23), \text{ for } k\gg1, \label{eq:binding:energy}
\end{equation}
where, $k_BT$ is the Boltzmann factor.
The factor of $\frac32$ is added in hind-foresight, as it makes some formulas cleaner.
Equation \eqref{eq:binding:energy} is only true asymptotically for large clusters. 
For small clusters, the separation between `bulk' and `surface' is
artificial, and we do not expect \eqref{eq:binding:energy} to be
quantitatively accurate. 
In particular, $\eps_1=0$ since the binding energy is the change in the
energy from the unbound state and a cluster with one particle is
unbound.

Next we consider the \emph{free energy} costs to create a $k$-cluster from the monomer bath.
The bath is characterized by the density $\rho_1$ of monomer, measured in units of $\oneover{v}$, where $v$ is the volume of each monomer.
In other words, $\rho_1$  is the volume fraction occupied by monomers

The free energy cost to create a $k$-cluster from the monomer bath is 
\begin{align}
  g_k &= -\eps_k -k_BT k \log \rho_1\notag\\
&\sim -k_BT\brk{(\alpha+\log \rho_1)k-\frac32\sig k^\frac23},
  \text{ for } k\gg 1.\label{eq:free:energy}
\intertext{
Here, $k_B T \log \rho_1$ is the chemical potential of a monomer in
the bath.
Rewriting the $k\gg1$ asymptotic form of the free energy gives insight
into the existence of a critical monomer density, $\rho_s \equiv
e^{-\alpha}$. 
We call $\rho_s$ the \emph{saturation density} of monomers.
Setting $\alpha=\log \oneover{\rho_s}$ in \eqref{eq:free:energy} gives}
  g_k&\sim k_BT\brk{\frac32\sig k^\frac23-k \log \frac{\rho_1}{\rho_s}}.\notag
\end{align}
Thus, when $\rho_1<\rho_s$ the free energy increases with $k$, allowing for an equilibrium. 
When $\rho_1>\rho_s$, the free energy attains its maximum at the critical cluster size $k_c$,

\begin{equation}
  k_c\sim \brk{\frac{\sig}{\log \frac{\rho_1}{\rho_s}}}^3, \text{
  as } \rho_1\goto\rho_s^+.\label{eq:critical:k}
\end{equation}
We  investigate the implications of this critical value later.
\subsection{Kinetics and Detailed Balance}

\begin{figure}[h!]
\centerline{\resizebox{6cm}{!}{\includegraphics{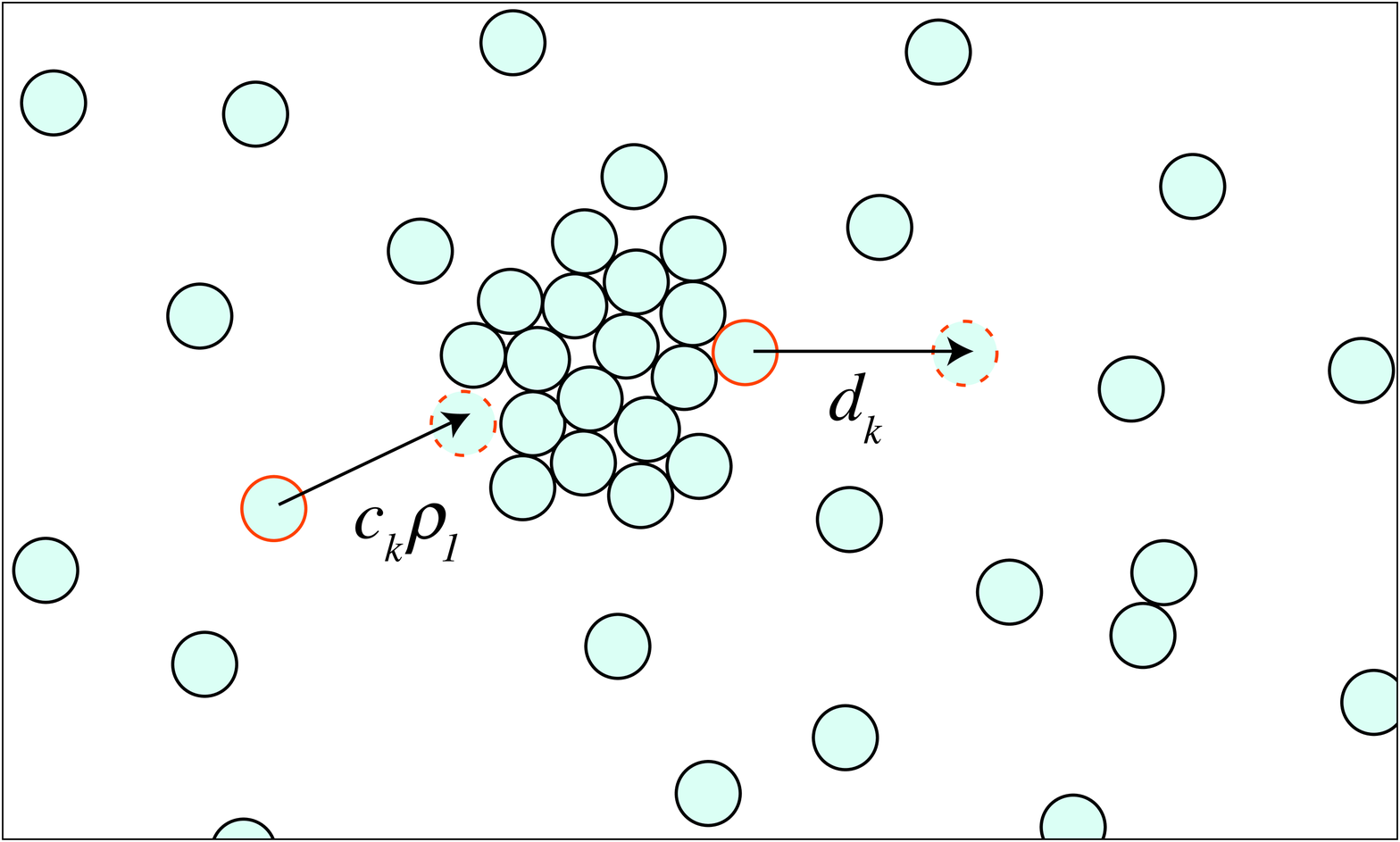}}}
\caption[Surface activated growth processes]{The exchange of particles between a cluster and the monomer
  bath. $c_k\rho_1$ is the rate at which a monomer gets added to the
  cluster, and $d_k$ is the rate at which monomers leave.}
\end{figure}

To study the rate of change in the cluster size we need to model the allowed reactions. 
BD allows two types of reactions:(1) A monomer in the bath can join the cluster; (2) A particle on the surface of the cluster can  dissociate from it and enter the bath. 
Mathematically, the two reactions are modeled as independent Poisson processes.
Let us call the rate at which particles leave a $(k+1)$-cluster $d_k$, and
the rate at which particles join a $k$-cluster $c_k\rho_1$.
We work under the assumption that the solution is dilute, that is,
$\rho_1\ll 1$.
It is reasonable to expect that in the dilute limit, the growth
rate is proportional to $\rho_1$.
Thus, we include $\rho_1$ in the growth rate, $c_k\rho_1$, so that both $c_k$ and
$d_k$ are independent of $\rho_1$. 

Initially we assume a uniform $\rho_1$, so it is a global parameter. 
When we add diffusion, we are a little more careful and take $\rho_1$
to be the density at the surface of the cluster. 

Since the dissociation happens on the boundary of the cluster, we
expect $d_k$ to be proportional to the surface area, which, in turn,
is proportional to $k^{2/3}$,
\begin{equation}
  d_k = \omega k^\frac23.\label{eq:d_k}
\end{equation}
Here, $\omega$ includes a per-particle  dissociation rate and a geometric factor.

A standard \emph{detailed balance} argument relates $c_k$ to $d_k$.
Let $\rho_1$ be the value of monomer density so that a $k$-cluster is in
equilibrium with the monomer bath. 
That is, the cluster has no net tendency to grow nor shrink, hence, the adsorption rate, $\rho_1c_k$, should match the emission rate, $d_k$.
The physical requirement for the equilibrium is that the free
energy is unchanged when a monomer is taken from the bath and added to the cluster,
\begin{equation*}
  g_{k+1}-g_k=-\eps_{k+1} + \eps_k -k_BT\log\rho_1 =0,
\end{equation*}
hence,
\begin{equation*}
  \rho_1 = e^{\frac{\eps_{k} - \eps_{k+1}}{k_BT}}.
\end{equation*}
The detailed balance relation between $c_k$ and $d_k$ is therefore,
\begin{equation}
  c_k = e^{\frac{\eps_{k+1} - \eps_{k}}{k_BT}}d_k. \label{eq:detailed:balance}
\end{equation}
We  use the adsorption and emission rates, $c_k\rho_1$ and $d_k$, to
derive a kinetic equation for the expected change in size of a cluster. 
In a small time span $\delta t$, the expected  change in cluster size,
$\delta k$, is:
\begin{equation}
  \Average{\delta k} = ( c_k\rho_1 - d_k) \delta t.\label{eq:expected:size:change}
\end{equation}
Using the model for the binding energy of large clusters \eqref{eq:binding:energy}
and  relation \eqref{eq:detailed:balance} between $c_k$ and $d_k$,
equation \eqref{eq:expected:size:change} can be rewritten as
\begin{equation*}
  \Average{\delta k} \sim \omega \brk{
 \frac{\rho_1-\rho_s}{\rho_s} k^\frac23-\sig k^\frac13 } \delta t, \text{ for } k\gg 1.
\end{equation*}
In the appendix we show that
if the variance in $k$ is much smaller than $\Average{k}$ initially,
it will remain so, as long as $\Average{k}$ is bounded away from the
critical size $k_c$ in \eqref{eq:critical:k}.
In this case we approximate the evolution of $k(t)$ for a given
cluster as deterministic, and governed by the ODE
\begin{equation}
  \dot{k} = \omega \brk{\eta k^\frac23 - \sig k^\frac13 }.\label{eq:BD:kinetics}
\end{equation}
Here, $\eta$ is the \emph{super-saturation}, defined by
\begin{equation}
  \eta = \frac{\rho_1-\rho_s}{\rho_s}.\label{eq:supersaturation}
\end{equation}
The super-saturation in \eqref{eq:supersaturation} is generally a function of time, due to the exchange of particles between clusters and the monomer bath.
By conservation of the \emph{total} particle density, the average value
of $\rho_1$, and hence $\eta$, is determined from the densities of all
clusters with $k\ge2$. 
Within the framework of BD, which assumes that the monomer density is uniform, we simply
set $\eta$ to this average value, and, in this sense,
\eqref{eq:BD:kinetics} is the BD prediction for cluster growth.
However, if the diffusivity of monomers is finite, the density of
monomers seen at the surface of the cluster will be different from the
average value far away.
We propose that \eqref{eq:BD:kinetics} holds generally, with $\eta$
equal to the value of super-saturation seen at the \emph{surface} of
the cluster.
Equations \eqref{eq:BD:kinetics} and \eqref{eq:supersaturation} expose
the \emph{kinetic} significance of the saturation density
$\rho_s= e^{-\alpha}$, and the critical cluster size $k_c$ in
\eqref{eq:critical:k}, which we rewrite using $\eta$,
\begin{equation}
  k_c\sim
  \brk{\frac{\sig}{\eta}}^3, \text{ as } \eta\goto 0^+.\label{eq:critical:k:2}
\end{equation}
If the surface value of $\rho_1$ is less than $\rho_s$, all the clusters
shrink, regardless of $k$.
For $\rho_1>\rho_s$, i.e. $\eta>0$, the critical size, $k_c$, separates growing,
\emph{super-critical} clusters ($k>k_c$) from shrinking, \emph{sub-critical} ones ($k<k_c$).

It should be noted that while the expected change in size of sub-critical clusters is negative, there is a small probability for a sub-critical cluster to grow and become super-critical.
The rate at which this happens is estimated by the Zeldovich
formula, to which  we come back in section \ref{sec:zeldovich}.
First we add finite diffusion of monomers to the model and see how it
affects the growth rate.

\begin{figure}[h!]
  \centerline{
\resizebox{6cm}{!}{\includegraphics{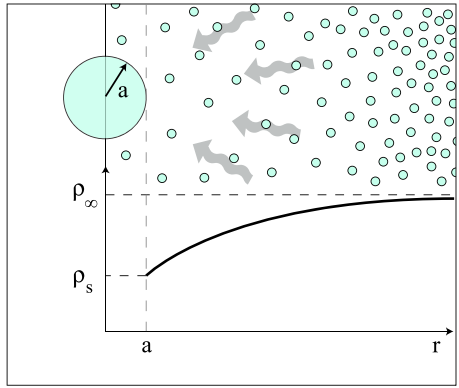}}}
\caption[Diffusion driven growth]{The cluster is surrounded by a \emph{inhomogeneous} monomer
  bath. The flux of monomers in the bath is diffusive, and the cluster
  reacts to the \emph{local} monomer density.}
\end{figure}

\section{Adding Monomer Diffusion}
Diffusion is the usual model of transport for monomers in the bath.
For a finite diffusion coefficient, $D$, the exchange of particles
between a cluster and the bath directly outside of it leads to
non-uniform monomer density.
Hence, the monomer concentration, $\rho_1$, is a function of position and 
time that satisfies the diffusion PDE.

To add diffusion to the BD model, we make a couple of additional
assumptions. 
The cluster is taken to be spherical, filled with monomers, each
taking volume $\nu$.
Thus, the number of particles in the cluster, $k$, and the radius of
the cluster, $a$, are related by
\begin{equation}
  k\nu = \frac{4\pi}{3}a^3.\label{eq:cluster:radius}
\end{equation}
The monomer density is assumed to be radially symmetric (with the
cluster centered at the origin.) 
Thus, the density, $\rho_1(r,t)$, satisfies the radially symmetric diffusion PDE in $\R^3$
\begin{align}
   \partial_t \rho_1 &= \frac{D}{r^2} \partial_{r}(r^2 \partial_r \rho_1) & &\text{ in } r>a,\label{eq:diffusion}
\end{align}

There are two BC at $r=a$: one is the kinetics due to BD as in equation
\eqref{eq:BD:kinetics}.
That is, $\rho_1(a,t)$ is related to $k$ and $\dot k$ by
\begin{equation}
  \dot{k} = \omega \brk{\eta(a,t) k^\frac23 - \sig k^\frac13 }\label{eq:BD:kinetics:2},
\end{equation}
where, 
\begin{equation*}
  \eta(a,t) = \frac{\rho_1(a,t)-\rho_s}{\rho_s}
\end{equation*}
is the super-saturation seen at the surface of the cluster.

The second BC results from conservation of particles.
The particles that are added to the cluster can come from two possible
sources: particles in the solution surrounding the cluster, which join
the cluster as it engulfs them, and  particles added by the diffusive flux.
This can be put in a simple equation:
\begin{equation}
  \dot a ( 1-\rho_1(a,t)) = D (\partial_r \rho_1)(a,t)=D\rho_s(\partial_r\eta)(a,t).\label{eq:FBP}
\end{equation}

Finally, there is a BC at $\infty$: 
the monomer concentration has the asymptotically uniform value
$\rho_\infty$ far from the clusters,
\begin{equation}
    \rho_1\goto\rho_\infty \text{ as } r\goto\infty.\label{eq:far:field}
\end{equation}
In the context of the full aggregation problem, $\rho_\infty$ is
generally a function of time, which  follows from overall conservation
of particles.
Equations (\ref{eq:cluster:radius}--\ref{eq:far:field}) constitute a
free boundary problem (FBP) for $a(t)$ and $\rho_1(r,t)$ in $r>a$.

The analysis of the FBP begins by identifying suitable non-dimensional
variables.
We assume that the local super-saturation
\begin{equation*}
  \eta(r,t)\equiv \frac{\rho_1(r,t)-\rho_s}{\rho_s}.\label{eq:super:saturation}
\end{equation*}
is uniformly small in $r>a$.
Hence, we introduce $\eps$ as a gauge parameter for $\eta(r,t)$ and we
replace $\eta(r,t)$ in (\ref{eq:cluster:radius}--\ref{eq:far:field})
by $\eps\eta(r,t)$.
The analysis of the FBP is based on an $\eps\goto0$ limit process.
However, we do not yet take the $\eps\goto0$ limit. 
It remains to determine the scaling of the other variables $k,r,a$ and
$t$ with $\eps$.
These follow from simple physical balances: 
in the `kinetic' BC \eqref{eq:BD:kinetics:2}, $\eta(r,t)$ has order of
magnitude $\eps$, so the two terms of the RHS balance when $k$ has
magnitude $\brk{\frac{\sig}{\eps}}^3.$
Notice that this is the critical cluster size \eqref{eq:critical:k:2}
for $\eta=\eps$.
The scalings of $\eta$ and $k$ are recorded in the scaling table
\begin{center}
Scaling Table\\
\begin{tabular}{l*{4}{>{$}c<{$}}}
Variable&\eta&k&a,r&t\\
\hline\\[-4mm]
Unit&\eps&\brk{\frac{\sig}{\eps}}^3&\frac{\sig}{\eps}\nu^{1/3}&\frac{\sig}{\omega\eps^2}
\end{tabular}
\end{center}
The unit of $r$ and $a$ is the radius of a cluster with
$k=\brk{\frac{\sig}{\eps}}^3$ particles.
The unit of $t$ is chosen to balance the LHS of the BC
\eqref{eq:BD:kinetics:2} with the two terms on the RHS.

The equations of the non-dimensional FBP that follow from
(\ref{eq:cluster:radius}--\ref{eq:far:field}) are 
\begin{align}
  k&=\frac{4\pi}{3}a^3,\label{eq:cluster:radius:ND}\\
  \partial_t \eta&=\frac{D}{\omega \nu^{2/3}\sig} \oneover{r^2}\partial_{r}(r^2\partial_r\eta), \text{ in } r>a, \label{eq:diffusion:ND} \\
    \dot k &= \eta(a,t)  k^{2/3} -k^{1/3},\label{eq:BD:scaled}\\
  \dot a \brk{\rule{0cm}{4mm}1-\rho_s(1+\eps\eta(a,t))} & = \frac{D}{\omega \nu^{2/3}}
  \frac{\eps\rho_s}{\sig}  (\partial_r\eta)(a,t)\label{eq:conservation:ND}\\
\eta&\goto\eta_\infty \text{ as } r\goto\infty. \label{eq:eta:infty}
\end{align}
Here, $\eta_\infty$ is the asymptotically uniform value of $\eta(r,t)$
as $r\goto\infty$.
The characteristic time of its variation is assumed to be comparable
to or larger than the unit of time $\frac{\sig}{\omega\eps^2}$ from the
scaling table.

The dimensionless constant $\frac{D}{\omega \nu^{2/3}}$ can be interpreted as a ratio of characteristic times for two different physical processes.
Recall that $\oneover{\omega}$ is the characteristic time for a monomer on the surface of a cluster to dissociate into the solution. 
The ratio $\frac{\nu^{2/3}}{D}$ is the characteristic time for a monomer to diffuse a distance comparable to its own size.
The conventional assumption is that the ``dissociation time'' is much longer than the ``diffusion time,'' so that 
\begin{equation*}
  \oneover{\omega}\gg \frac{\nu^{2/3}}{D} \quad \Longleftrightarrow \quad \frac{D}{\omega \nu^{2/3}}\gg1. 
\end{equation*}
In this limit, the diffusion equation \eqref{eq:diffusion:ND} reduces to a radial Laplace equation
\begin{equation*}
  \partial_{r}(r^2\partial_r\eta)=0.
\end{equation*}
The solution with  $\eta=\eta_\infty$ as $r\goto\infty$ is
\begin{equation}
  \eta(r,t) = \eta_\infty + (\eta(a,t)-\eta_\infty)\frac ar.\label{eq:eta:lap}
\end{equation}
The time dependence of $\eta$ is implicit due to the time-dependence
of its values at $r=a$ and $r=\infty$.

In the dilute limit, with $\rho_s\ll1$, the conservation equation \eqref{eq:conservation:ND} reduces to
\begin{equation*}
  \dot a = \frac{D}{\omega \nu^{2/3}} \frac{\eps \rho_s}{\sig} (\partial_r \eta)(a,t).
\end{equation*}
Substituting $\eta$ from \eqref{eq:eta:lap}, this is becomes
 \begin{equation}
  a \dot a = \frac{D}{\omega \nu^{2/3}} \frac{\eps \rho_s}{\sig} \brk{\eta_\infty-\eta(a,t)\rule{0mm}{4mm}}.\label{eq:a:dot:diff}
\end{equation}
Equation \eqref{eq:a:dot:diff} can be converted into an equation for
$\dot k$ using \eqref{eq:cluster:radius:ND}. 
This gives
\begin{equation}
  \dot k = \eps\mu(\eta_\infty-\eta(a,t))k^\third, \label{eq:k:dot:diff}
\end{equation}
where $\mu$ is the dimensionless constant:
\begin{equation*}
  \mu = (3 \cdot 16 \pi^2)^\third \brk{\frac{D}{\omega\nu^{2/3}}}\brk{ \frac{\rho_s}{\sig}}.
\end{equation*}
Since $\mu$ is a product of a large number, $\frac{D}{\omega \nu^{2/3}}$,
and a small number, $\frac{\rho_s}{\sig}$, it can be large or small. 
We therefore entertain any value of $\mu$.

The two equations for $\dot k$, \eqref{eq:BD:scaled} and
\eqref{eq:k:dot:diff},  involve the super-saturation at the surface of the cluster, $\eta(a,t)$. 
Solving for  $\dot k$ and $\eta(a,t)$ gives
\begin{align}
  \dot k &= \frac{\eta_\infty k^{2/3}-k^{1/3}}{1+\frac{k^{1/3}}{\eps\mu}}, \label{eq:new:cluster:kinetics}\\
  \eta(a,t) &= \frac{\eta_\infty + \oneover{\eps\mu}}{1 + \frac{k^{1/3}}{\eps\mu}}.\label{eq:eta_s}
\end{align}
ODE \eqref{eq:new:cluster:kinetics} indicates a \emph{second} characteristic cluster size besides $k_c$. 
In units of $\brk{\frac{\sig}{\eps}}^3$ this cluster size is
  \begin{equation*}
    k_* = (\eps\mu)^3.
  \end{equation*}
In the original variables, $k_*$ is a combination of basic physical constants:
\begin{equation}
  k_* = (\sig\mu)^3=(3 \cdot 16 \pi^2) \frac{D^3\rho_s^3}{\omega^3 \nu^2}.\label{eq:k_*}
\end{equation}
Notice that for $k\ll k_*$  equation \eqref{eq:new:cluster:kinetics}
asymptotes to
BD\@.
For $k\gg k_*$, the asymptotic form of
\eqref{eq:new:cluster:kinetics} is
\begin{equation}
  \dot k \sim \eps\mu\brk{\eta_\infty k^\third-1}.\label{eq:LS}
\end{equation}
Restoring original units, \eqref{eq:LS} becomes 
\begin{equation}
  \dot k = d \brk{\eta_\infty k^\third-\sig},\quad d = (3\cdot
  16\pi^2)^\third\frac{D\rho_s}{\nu^\frac23}, \label{eq:LS:dim}
\end{equation}
which is the standard result for diffusion limited growth (DLG)
\cite{LS61}.
Equation \eqref{eq:eta_s} shows how the surface value of super-saturation differs from the uniform value, $\eta_\infty$, far from the cluster. 
Notice that it is a function of $k$. 
We convert it into an equation for $s_k$, the value of monomer density
seen at the surface of a $k$-cluster (again, in original units):
\begin{equation}
  s_k=\rho_s\brk{1+\frac{\eta_\infty+\oneover{\mu}}{1+\frac{k^{1/3}}{\sig\mu}}}.\label{eq:s_k}
\end{equation}
This will be important, when we examine the whole ensemble of clusters, and formulate evolution equations for cluster densities.
\subsection{Physical Meaning of $k_*$}
We show that $k_*$ is the characteristic size of clusters for which finite diffusion induces a significant relative difference between $\eta_\infty$ and $\eta(a,t)$.
That is, $\delta\eta\equiv\eta_\infty-\eta(a,t)$ is comparable in magnitude to $\eta_\infty$.
A simple examination of two physical balances is sufficient. 
First, the cluster's growth rate balances the diffusive influx of monomers.
This is expressed by
\begin{equation}
  \dot k = a^2 D \frac{\brk{\frac{\rho_s}{\nu}}\delta\eta}{a}.\label{eq:diffusion:mag}
\end{equation}
Here, the equality means `order of magnitude balance'. 
In the RHS, $\frac{\rho_s}{\nu}\delta\eta$ is the difference between monomer densities at $\infty$, and on the surface, expressed in the conventional unit of 1/volume. 
For quasi-static diffusion, the diffusion zone about the cluster of radius $a$ has thickness $a$, so $\frac{\brk{\frac{\rho_s}{\nu}}\delta\eta}{a}$ estimates $(\partial_r\rho_1)(a,t)$ and the influx of monomers per unit area into cluster is estimated by multiplying this by $D$. 
Finally, multiplying by the area, proportional to $a^2$, gives the cluster growth rate, $\dot k$. 

Second, the magnitude of $\dot k$ as dictated by the surface reactions \eqref{eq:BD:kinetics:2} is 
\begin{equation}
  \dot k = \omega\eta k^\frac23.\label{eq:surface:reaction:mag}
\end{equation}
Enforcing the equivalence of \eqref{eq:diffusion:mag} and
\eqref{eq:surface:reaction:mag} and using $k=\nu a^3$ (order of magnitude equality again), we find
\begin{equation*}
  \frac{\delta\eta}{\eta}=\frac{\omega \nu^{2/3}}{D\rho_s}k^\third=\brk{\frac{k}{k_*}}^\third.
\end{equation*}
We see that $\delta\eta$ is comparable to $\eta$ when $k$ is comparable to $k_*$.

\subsection{Critique of DLG and its `paradox'}
We briefly examine the `traditional' derivation of ODE \eqref{eq:LS} for DLG, within the framework of the non-dimensional free boundary problem (\ref{eq:cluster:radius:ND}--\ref{eq:eta:infty}).
Given $\eta(a,t)$, equation \eqref{eq:k:dot:diff} gives the growth rate
of the cluster that follows from diffusive flux of monomers.
In the traditional analysis of DLG, $\eta(a,t)$ is chosen so that the
cluster is in equilibrium with the monomer bath that surrounds it.
Under the current non-dimensionalization, this `critical nucleus' BC reads
\begin{equation}
  \eta(a,t) = k^{-\third}\label{eq:critical:ND}.
\end{equation}
Substituting \eqref{eq:critical:ND} for $\eta(a,t)$ in
\eqref{eq:k:dot:diff} leads directly to the ODE \eqref{eq:LS}.

By inspection, we see that the traditional equations
(\ref{eq:LS}, \ref{eq:critical:ND}) arise by taking the $\eps\mu\goto0$
limit of (\ref{eq:new:cluster:kinetics}, \ref{eq:eta_s})  with $k$
fixed.
The alternative limit process, $\frac{k}{k_*}\goto\infty$ with
$k_*=\eps\mu$ fixed is more physical: the value of $\eps\mu$ is set by
material properties and initial conditions, and we expect that
clusters eventually grow to sizes $k\gg k_*$. 
We have already seen that the ODE \eqref{eq:new:cluster:kinetics} for
$k$ converges to the DLG result in this limit, but the expression
\eqref{eq:eta_s} for the surface value of monomer density does not
converge to the DLG boundary condition \eqref{eq:critical:ND}.
Instead, 
\begin{equation}
  \eta(a,t) \sim (1+\mu \eta_\infty)k^{-\third},\label{eq:surface:mono:den}
\end{equation}
which has an additive term $\mu\eta_\infty$ in the prefactor of
$k^{-1/3}$ not present in $\eqref{eq:critical:ND}$.
A mathematical critique of the `critical nucleus' boundary condition
\eqref{eq:critical:ND} is simple; it results from formally neglecting
$\dot k$ in the LHS of the `surface kinetics' boundary condition
\eqref{eq:BD:scaled}.
In our result, $\dot k$ balances the RHS, even in the limit
$\frac{k}{k_*}\goto\infty$, resulting in \eqref{eq:surface:mono:den}.

Recall that the `critical nucleus' boundary condition in
traditional DLG looks paradoxical because the `cluster sits on top of
a free energy maximum'.
A `lazy' deconstruction might say: ``Nothing to explain, the critical
nucleus boundary condition is simply incorrect in the (more physical)
limit $\frac{k}{k_*}\goto\infty$ with $k_*$ fixed.'' 
Another easy explanation  looks at the free energy. The free energy
\eqref{eq:free:energy} refers to a simple cluster surrounded monomers of
\emph{uniform}  density, whereas the actual kinetics we consider
involves a \emph{non-uniform} density $\rho_1(r,t)$ in $r>a$.
The actual free energy takes into account the functional dependence of
$\rho_1(r,t)$ in $r>a$.
These remarks indicate that the `paradox' in its original form is
na\"ive.
Nevertheless, it points to some physics that is not expressed in the
quasi-static model (\ref{eq:new:cluster:kinetics}, \ref{eq:eta_s}) of cluster growth as it stands.

Suppose that we place a cluster of size $k$ into a uniform monomer bath
that has the `wrong` monomer density, not equal to the surface value
$s_k$ given in \eqref{eq:s_k}.
In order for our model to be
plausible, the surface value of $\rho_1$ should rapidly relax to $s_k$
in  \eqref{eq:s_k}. 
We now show that the full free boundary problem
(\ref{eq:cluster:radius:ND}--\ref{eq:eta:infty}) implies such a
relaxation transient.

\subsection{The Stability of the Free Boundary Problem}
The relaxation transient is characterized by a balance of the time and
space derivatives in the diffusion PDE \eqref{eq:diffusion}.
Hence the characteristic time of the relaxation transient is
\begin{equation}
  t_r \equiv \frac{a^2}{D},\label{eq:relax:time}
\end{equation}
where $a$ is the cluster radius.
The relative change of the cluster radius in this characteristic time
is small: from \eqref{eq:FBP}, $\dot a$ has the order of magnitude
$\frac{D\rho_s\eps}{a}$.
Hence, the relative change of radius in time $t_r$ has magnitude
$\eps\rho_s$.
The small relative change in cluster radius means that the cluster
radius is asymptotically constant during the relaxation transient, and
it remains to derive from the full free boundary problem
(\ref{eq:cluster:radius}--\ref{eq:far:field}) a reduced boundary value
problem for $\eta(r,t)$ in $r>a$, with $a$ fixed.
We use the previous units in the scaling table for all variables
except time $t$. 
For $t$ we use $t_r$ in \eqref{eq:relax:time} with 
$a$ replaced by the characteristic cluster radius,
$\frac{\sig}{\eps}\nu^{1/3}$.
The reduced boundary value problem is
\begin{align}
\partial_t \eta = \oneover{r^2} \partial_r(r^2 \partial_r \eta)\text{
  in } r>a,\label{eq:trans:PDE}\\
\lambda(\partial_r\eta)(a,t) = \eta(a,t) -
  k^{-\third},\label{eq:trans:BC}\\
\eta(r,t)\goto\eta_\infty\text{ as } r\goto\infty,\label{eq:trans:infty}
\end{align}
in the limit $\eps\goto0,$ and $\lambda \equiv\brk{\frac{3}{4\pi}}^\third
  \eps \mu$ fixed.
The far-field super-saturation, $\eta_\infty$, is assumed to vary on a
characteristic time much longer than $t_r$, so $\eta_\infty$ is
effectively constant.

The time-independent solution of
(\ref{eq:trans:PDE}--\ref{eq:trans:infty}) for $\eta(r,t)$ is
\eqref{eq:eta:lap} with $\eta$ on $r=a$ given by \eqref{eq:eta_s}.
We show that this time-independent solution is asymptotically stable.
We notice that
\begin{equation}
  E\equiv\frac{\lambda}{2}\int_a^\infty r^2(\partial_r\eta)^2  \, dr +
  \frac{a^2}{2} \brk{\eta(a,t)-k^{-\third}}^2\label{eq:Lyapunov}
\end{equation}
is a Lyapunov functional for equations
(\ref{eq:trans:PDE}--\ref{eq:trans:infty}).
The time derivative of $E$,
\begin{equation*}
  \dot E = -\lambda \int_a^\infty r^2\eta_t^2 \,dr,\label{eq:Lyapunov:deriv}
\end{equation*}
is found by time-differentiation of \eqref{eq:Lyapunov}, integration by parts, and use
of the PDE \eqref{eq:trans:PDE} and BC \eqref{eq:trans:BC}.
Since $E$ is positive definite, and $\dot E\le0$, it follows that $\eta(r,t)$ converges
to the time-independent solution. 
We conclude that if the surface monomer concentration is initially
different from the quasi-static value \eqref{eq:eta_s}, it relaxes to
it in characteristic time $t_r$.

\section{Evolving Distribution of Cluster Sizes}
\label{sec:zeldovich}
The kinetics equation \eqref{eq:new:cluster:kinetics} requires $\eta_\infty$, the super-saturation far from any cluster.
To find $\eta_\infty$ we look at the joint evolution of all the clusters, each assumed to follow the dynamics in \eqref{eq:new:cluster:kinetics}.
The clusters are coupled by the combined effect they have on the
monomer density, and consequently, on the super-saturation.

Let $\rho_k(t)$ be the average spatial density (in units of $\oneover{\nu}$) of $k-$clusters at time $t$. 
Assuming that the total particle density has a fixed value, $\rho$,
the $\rho_k$  satisfy a particle conservation equation. 
Since a $k$-cluster is made of $k$ particles, the total particle density, $\rho$, must satisfy
\begin{equation}
  \rho = \sum_{k=1}^\infty k \rho_k.\label{eq:conservation}
\end{equation}
With the help of \eqref{eq:conservation}, the space averaged super-saturation can be rewritten as a function of the cluster densities $\rho_k$ with $k\ge2$:
\begin{equation}
  \frac{\rho_1-\rho_s}{\rho_s} = \frac{\rho-\rho_s}{\rho_s} -\oneover{\rho_s}\sum_{k=2}^\infty k\rho_k.\label{eq:rho_1:con}
\end{equation}
In the dilute limit with inter-cluster distances much greater than cluster radii, we expect that the super-saturation is asymptotically uniform, throughout most of the monomer bath far from clusters.
In this case, that asymptotically uniform value, $\eta_\infty(t)$, should be well approximated by the spatial average \eqref{eq:rho_1:con},
\begin{equation}
  \eta_\infty=\frac{\rho-\rho_s}{\rho_s} -\oneover{\rho_s}\sum_{k=2}^\infty k\rho_k.\label{eq:super-sat:con}
\end{equation}

We turn to the  evolution of the densities.
The $\rho_k$ obey kinetic equations associated with the
reactions
\begin{equation*}
  k \text{-cluster} + \text{monomer} \rightleftarrows (k+1)\text{-cluster}.
\end{equation*}
The equations are
\begin{align}
  \dot{\rho_k}&=j_{k-1}-j_k,\label{eq:discrete:kinetic}
\intertext{for $k\ge2$, where the discrete flux $j_k$ is  the \emph{net} rate of creation of a $(k+1)$-cluster from a
$k$-cluster,}
j_k&\equiv c_ks_k\rho_k -d_k\rho_{k+1}.\label{eq:discrete:flux}
\end{align}
As before, $d_k$ is the rate constant for shedding a monomer from the
surface of a $(k+1)$-cluster.
For $k\gg1$ it has the asymptotic behavior \eqref{eq:d_k}, proportional to surface area.
The prefactor $c_ks_k$ of $\rho_k$ in \eqref{eq:discrete:flux} is the rate constant for \emph{adding} a monomer.
Recall that $s_k$ is the value of monomer density \emph{seen at the surface} of a $k$-cluster, given by \eqref{eq:s_k}, and $c_k$ is related to $d_k$ by the detailed balance condition \eqref{eq:detailed:balance}.
A more explicit formula for $j_k$ displaying the $k$-dependence of surface monomer concentration and detailed balance is
\begin{equation}
j_k =   d_k\brk{e^{\frac{\eps_{k+1}-\eps_k}{k_BT}}s_k\rho_k-\rho_{k+1}}.\label{eq:j_k}
\end{equation}
Equations (\ref{eq:discrete:kinetic}, \ref{eq:j_k}) can be summarized as discrete advection-diffusion equations,
\begin{equation}
  \dot\rho_k+D^-\Brk{d_k\brk{1-e^{\frac{\eps_{k+1}-\eps_{k}}{k_BT}}s_k}\rho_k-d_kD^+\rho_k}=0 \text{ for } k\ge2.\label{eq:discrete:ad:dif}
\end{equation}
Here $D^+, D^-$ are, respectively, the forward and backward difference operators.
In these equations, the surface monomer density $s_k$ contains the super-saturation $\eta_\infty$ as a parameter, and $\eta_\infty$ is connected to the $\rho_k$ according to \eqref{eq:super-sat:con}.
So we see explicitly how the cluster densities are coupled to each other via the super-saturation.
The $k$-dependence of $s_k$ induced by the finite diffusivity of monomers is the essential difference from classical BD\@. 
We recover classical BD by taking $\mu\goto\infty$, which in turn results from $\frac{D}{\omega \nu^{2/3}}\goto\infty$. 
Then $s_k$ in \eqref{eq:s_k} reduces to $\rho_s(1+\eta_\infty)$, which is the \emph{uniform} value of $\rho_1$ assumed in classical BD.

\subsection{Equilibrium}
\emph{Equilibria} are time independent densities, $\tilde\rho_k$, so that all the fluxes $j_k$ are zero, and the sum \eqref{eq:conservation}, which gives the total particle density, is convergent.
Setting $j_k=0$ in \eqref{eq:j_k} gives a recursion relation that determines $\tilde\rho_k$ from $\rho_1$,
\begin{equation}
  \tilde\rho_k = \rho_1^ke^{\frac{\eps_k}{k_BT}}, \text{ for } k\ge2.\label{eq:rho_k:equil}
\end{equation}
Here, we used $s_k=\rho_1$ for all $k$, since $\rho_1$ should be uniform in the equilibrium case. 
Substituting these $\tilde\rho_k$ into \eqref{eq:conservation} gives,
\begin{equation}
  \rho = \rho(\rho_1)\equiv \sum_{k=1}^{\infty}k\rho_1^ke^{\frac{\eps_k}{k_BT}}.\label{eq:conservation:equil}
\end{equation}
Hence, equilibria exist for monomer densities $\rho_1$ so that this series converges.
For large values of $k$, the binding energy can be written as 
\begin{equation*}
  \eps_k \sim k_BT\brk{\alpha k - \frac32\sig k^{2/3}}
\end{equation*}
Therefore, convergence happens for 
\begin{equation*}
  \rho_1 \le e^{-\alpha}=\rho_s.
\end{equation*}
In other words, equilibria exist only if the super-saturation is non-positive.
The largest value of  total particle density for which there is equilibrium is obtained by setting $\rho_1=\rho_s$ in \eqref{eq:conservation:equil}.
We denote this \emph{critical particle density} by $\rho_c$, 
\begin{equation}
  \rho_c=\rho(\rho_s).\label{eq:rho_c}
\end{equation}
Since $\rho_s\ll1$, the first few terms of the series give a close approximation of $\rho_c$.
\subsection{Zeldovich Nucleation Rate}
For positive super-saturation $\eta$, there is the critical cluster size $k=k_c$, for which the free energy cost to assemble a $k$-cluster from dissociated monomers is maximized. 
In the small super-saturation limit $\eta\goto0^+$, and $k\gg1$, it
follows from \eqref{eq:free:energy} and \eqref{eq:supersaturation},
that
\begin{equation}
  g_k \sim k_BT\brk{\frac32 \sig k^\frac23 -\alpha \eta},\label{eq:free:energy:2}
\end{equation}
and that the free energy cost (in units of $k_BT$) of the critical cluster is asymptotic to
\begin{equation*}
  g\equiv \max_{k} g_k \sim \frac{\sig^3}{{2\eta^2}}.
\end{equation*}
For small super-saturation, the free energy cost is high, and an analogy with the famous Arrhenius rate suggests that super-critical nuclei with $k>k_c$ are produced at a rate proportional to the exponential $e^{-g}$.
Since this proposed creation rate is exponentially small as $\eta\goto
0$, one might expect that after some initial transient, quasi-static
but non-equilibrium values of $\rho_k$ are established for $k$ on the
order of $k_c$, in which the discrete fluxes $j_k$ in
\eqref{eq:discrete:flux} are asymptotically equal to a uniform value $j$.
This $j$ is the creation rate of super-critical nuclei, proportional to $e^{-g}$. 
These essential ideas of nucleation kinetics are set forth in a famous paper of Zeldovich, on the nucleation of vapor bubbles for under pressurized liquid \cite{ZELD43}.
His starting point is a discrete system of kinetic ODE's like BD, but he first passes to a PDE limit of the ODE's and calculates the nucleation rate from the PDE\@.
Here, we implement the essential Zeldovich ideas, but within the framework of the discrete BD ODE's.

We work in the limit $k_c\ll k_*$, so for $k$ on the order of $k_c$ there is negligible difference between the surface value, $s_k$, of monomer density in \eqref{eq:s_k}, and the uniform value, $\rho_1$, far from clusters.
We show this: in equation \eqref{eq:s_k} for $s_k$, we see that $s_k\sim\rho_s(1+\eta_\infty) = \rho_1$ if $\oneover{\mu}\ll\eta_\infty$ and $\frac{k^{1/3}}{\sig\mu}\ll1$.
For $k_c=\brk{\sig/\eta_\infty}^3$ in \eqref{eq:critical:k:2} and $k_*=(\sig\mu)^3$ in \eqref{eq:k_*}, we find $\frac{k_c}{k_*}=\oneover{(\mu\eta_\infty)^3}$.
So, $\frac{k_c}{k_*}\ll 1$ implies $\oneover{\mu} \ll \eta_\infty$.
Furthermore, for $k$ on the order of $k_c$, $\frac{k^{1/3}}{\sig\mu}$ is on the order of $\brk{\frac{k_c}{k_*}}^{1/3}\ll1$.

In \eqref{eq:j_k}, we replace $s_k$ by $\rho_1$, and $j_k$ by $j$, to obtain a recursion equation that determines the $\rho_k$ for $k\ge2$ from $\rho_1$.
We write it as
\begin{equation}
  \frac{\rho_k}{\tilde\rho_k}-\frac{\rho_{k+1}}{\tilde\rho_{k+1}} = \frac{j}{d_k} e^{\frac{g_{k+1}}{k_BT}}.\label{eq:rho_k:quasi:stable}
\end{equation}
Here, $g_k$ is the free energy cost of a $k$-cluster, given in \eqref{eq:free:energy:2}, and $\tilde\rho_k$ denotes the solution \eqref{eq:rho_k:equil} of the homogeneous recursion relation with $j=0$, and $\tilde\rho_1=\rho_1$.
For positive super-saturation $\eta$, $\tilde\rho_k\goto\infty$ as $k\goto\infty$, and we expect that the $\rho_k$ in \eqref{eq:rho_k:quasi:stable} have $\frac{\rho_k}{\tilde\rho_k}\goto0$ as $k\goto\infty$.
Summing \eqref{eq:rho_k:quasi:stable} over $k$ gives a formula for $j$.
On the LHS, we get a telescoping sum with value 
\begin{equation*}
  \frac{\rho_1}{\tilde\rho_1}-\lim_{k\goto\infty}\frac{\rho_k}{\tilde\rho_k}=1-0=1.
\end{equation*}
Hence,
\begin{equation}
  1 = j\sum_{k=2}^{\infty}\oneover{d_{k-1}}e^{\frac{g_k}{k_BT}}.\label{eq:flux:sum}
\end{equation}
In the RHS, $g_k$ decreases linearly with $k$ as $k\goto\infty$, so the series converges.
In the small super-saturation limit $\eta\goto0^+$, we expect that the
sum on the RHS is dominated by terms  with $k$ near
$k_c\sim\brk{\frac{\sig}{\eta}}^3$, where $g_k$ attains its
maximum.
The relevant approximation to $g_k$ as $\eta\goto0^+$ and $k$ is on the order of $k_c$ is given by \eqref{eq:free:energy:2}.
Also, $d_k\sim \omega k^{2/3}$ as in \eqref{eq:d_k}.
Hence, \eqref{eq:flux:sum} has the asymptotic approximation 
\begin{equation}
  1 = \frac{j}{\omega}\sum_{k=2}^{\infty}k^{-\frac23}e^{\frac32\sig k^\frac23 - k\eta}.\label{eq:flux:sum:approx}
\end{equation}
The final step is the approximation of the sum by an integral, and evaluation of the $\eta\goto0^+$ limit by the saddle point method.
This leads to the approximation of $j$,
\begin{equation}
  j\sim\omega\sqrt{\frac{\sig}{6\pi}} e^{-\frac{\sig^3}{2\eta^2}}.\label{eq:zeldovich:rate}
\end{equation}
\section{Advection Signaling Problem}\label{sec:signaling:problem}

\begin{figure}[h!]
  \centerline{
\resizebox{8cm}{!}{\includegraphics{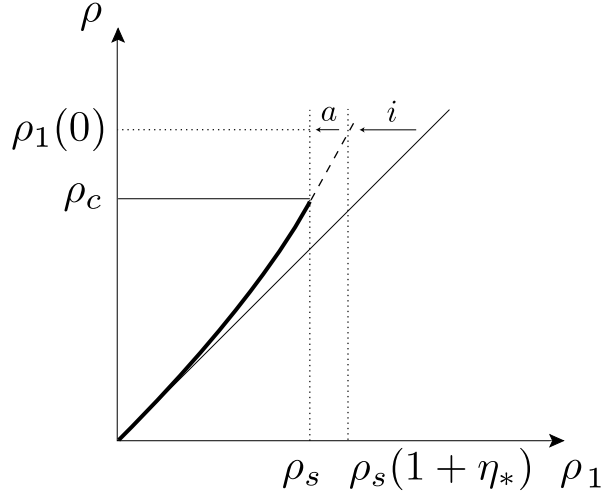}}}
\caption[The graph of $\rho(\rho_1)$ in
  $0\le\rho_1\le\rho_s$]{The darkened curve is the graph of $\rho(\rho_1)$ in
  $0\le\rho_1\le\rho_s$.
The dashed line is its linear interpolation into $\rho_a>\rho_s$.
It intersects $\rho=\rho_1(0)>\rho_c$ at $\rho_1=\rho_s(1+\eta_*)$.
The arrow labeled $i$ represents the decrease of $\rho_1$ during
  ignition, and the arrow $a$ represents the decrease during the subsequent
  aggregation process.}\label{fig:aggregation:process} 
\end{figure}

We examine the aggregation process, starting from a super-critical
density of particles, $\rho>\rho_c$, all in the form of monomers at
time $t=0$.
That is, 
\begin{equation*}
  \rho_1(0) = \rho>\rho_c,\quad \rho_k(0)=0 \text{ for } k\ge2.
\end{equation*}
There is an initial transient, called \emph{ignition}, in which the
first supercritical clusters appear. 
A detailed analysis of the ignition transient appears in a paper by
Bonilla {\it et al.} \cite{NBC05}.
Here we give a brief summary. First, sub-critical $(k<k_c)$ clusters are
created with quasi-static densities close to the values
$\tilde\rho_k$ in \eqref{eq:rho_k:equil}.
Of course, the value of $\rho_1$ becomes less than $\rho_1(0)=\rho$,
since these sub-critical clusters are created from monomers.
Hence, the appearance of the sub-critical quasi-static densities is
accomplished by the decrease of super-saturation from an initial value
of $\eta(0) = \frac{\rho-\rho_s}{\rho_s}$ to a smaller value, which we
denote by $\eta_*$. 
In appendix \ref{sub:sec:eta:eff}, we show that $\eta_*$ is related to $\rho-\rho_c$ by
\begin{equation}
  \rho-\rho_c \sim \eta_* \rho_s \rho'(\rho_s) \label{eq:eta:eff}
\end{equation}
as $\rho\goto\rho_c^+$.
Here, $\rho=\rho(\rho_1)$ is the equilibrium relation
\eqref{eq:conservation:equil} between $\rho_1$ and $\rho$ for
$0<\rho_1\le\rho_s$.
Figure~\ref{fig:aggregation:process} is a visualization of the
relation \eqref{eq:eta:eff} between $\eta_*$ and $\rho-\rho_c$.
The `completion' of the quasi-equilibrium in $1\le k< k_c=\brk{\frac{\sig}{\eta_*}}^3$ is
accompanied by the appearance of the first super-critical clusters
with $k>k_c$.
The rate of creation rises from zero to the Zeldovich rate in
\eqref{eq:zeldovich:rate} with $\eta=\eta_*$.
We assume $\eta_*$ is so that $k_c\ll
k_*$, so the Zeldovich rate indeed applies.

Now our focus shifts to the evolving distribution of the \emph{super-critical}
clusters.
The model we present here has three physical ingredients.
Two of them are the growth of the clusters, and their creation.
Both processes contain the super-saturation as a parameter, and the
remaining  ingredient is the connection of
super-saturation to the distribution of cluster sizes.

First, growth.
We expect a predominance of super-critical clusters with $k\gg k_*$ that undergo
diffusion limited growth.
Here is the heuristic argument for not resolving size scales
comparable to $k_*$ or smaller: once a cluster achieves super-critical
size, with $\frac{k-k_c}{k_c} \gg \eta^2$ (see the appendix \ref{sub:sec:det:BD})
it continues to grow  nearly deterministically, as shown in
section~3. 
Since the Zeldovich rate is exponentially small in $\eta$, an
exponentially long time elapses before the super-saturation shows an
significant decrease below the effective initial value, $\eta_*$,
established during ignition.
In this exponentially long time, we expect the super-critical clusters
to grow to sizes $k\gg k_*$, the regime of diffusion limited growth.

We assume that the characteristic cluster size $\bar k$ corresponding
to significant variations of the densities $\rho_k$ for $k\gg k_*\gg1$
is itself much larger than $k_*$, and this motivates a continuum
limit,
\begin{equation}
  \rho_k(t) \sim r(k,t).\label{eq:continuum}
\end{equation}
Here, $r(k,t)$ is a smooth function of its arguments, with the
characteristic scale $\bar k$ of $k$ much larger than $k_*$.
Substituting \eqref{eq:continuum} for $\rho_k$ into the discrete
advection-diffusion equations \eqref{eq:discrete:ad:dif} and using the
assumed largeness of $\bar k$, it follows that $r(k,t)$
asymptotically satisfies the advection PDE,
\begin{equation}
  \partial_t r + \partial_k (u\, r) =0. \label{eq:continuum:PDE}
\end{equation}
Here, the advection velocity $u=u(k,\eta)$ is identified from ODE
\eqref{eq:LS:dim} for  diffusion limited growth. 
We have,
\begin{equation}
 u(k,\eta)= d \brk{\eta \,k^\third -\sig},\label{eq:continuum:PDE:vel}
\end{equation}
where $\eta = \eta(t)$ is the `background' super-saturation, far from
any cluster.

Next, creation. 
In the analysis according to Zeldovich, recall that the discrete flux
$j_k$ in \eqref{eq:j_k} is asymptotically uniform for $k$ on the order
of $k_c$, with value $j$ given by \eqref{eq:zeldovich:rate}.
Here, we make the stronger assumption that the range of $k$ with the
asymptotically uniform value $j$ of $j_k$ extends to a scale of $k$
much larger than $k_*$ but smaller than the characteristic cluster
size $\bar k$ associated with the continuum limit
\eqref{eq:continuum}.
In this case, we expect an asymptotic matching between the continuum
limit of $j_k$, given by
\begin{equation}
  j_k \sim u(k,\eta) \, r(k,t),\label{eq:flux:BC}
\end{equation}
and the uniform value, $j$, of $j_k$, in some overlap domain of
cluster sizes $k$ much larger than $k_*$, but much smaller than $\bar
k$.
Since $k\gg k_*\gg k_c=\brk{\frac{\sig}{\eta_*}}^3$ in the overlap
domain, the dominant component of $u$ in \eqref{eq:continuum:PDE:vel}
is $d\,\eta\,k^{1/3}$ and \eqref{eq:flux:BC} reduces to
\begin{equation*}
  j_k \sim d\,\eta\,k^{1/3} r(k,t).
\end{equation*}
in the overlap domain.
Hence,  we propose the effective boundary condition (BC)
\begin{equation}
   d\,\eta\,k^{1/3} \, r(k,t) \goto j = \omega\sqrt{\frac{\sig}{6\pi}}e^{-\frac{\sig^3}{2\eta^2}}\quad \text{ as } k\goto0.\label{eq:continuum:BC}
\end{equation}
If $\eta = \eta(t)$ is known, the advection PDE
\eqref{eq:continuum:PDE} and creation BC \eqref{eq:continuum:BC}
lead to a simple determination of $r(k,t)$, starting from the initial
condition of pure monomer, $r(k,0)\equiv 0$ in $k>0$. 

It remains to connect $\eta(t)$ to $r(k,t)$.
In the particle conservation identity \eqref{eq:conservation},
we now distinguish between sub-critical and super-critical clusters sizes.
That is,
\begin{equation*}
  \rho =\sum_{\makebox[0mm]{$\scriptstyle 1\le k\le k_c$}} k \rho_k + \sum_{k_c < k}k \rho_k.
\end{equation*}
The sub-critical sum is approximated by substituting quasi-equilibrium
values \eqref{eq:rho_k:equil} for $\rho_k$ based on super-saturation
$\eta$,
\begin{equation*}
  \rho_k\sim\tilde\rho_k = (1+\eta)^k\rho_s^ke^{\frac{\eps_k}{k_BT}},
\end{equation*}
and then taking the limit $\eta\goto 0$. 
The details here are a re-run of the calculation in appendix \ref{sub:sec:eta:eff}. 
We get
\begin{equation*}
  \sum_{k=1}^{k_c} k\,\rho_k \sim \rho_c + \eta \rho_s \rho'(\rho_s).
\end{equation*}
The super-critical sum is approximated by the integral,
\begin{equation*}
  \int_0^{\makebox[0mm][l]{$\scriptstyle\, \infty$}} k\,r(k,t)\,dk.
\end{equation*}
Here, the lower limit is $k=0$ and not $k_c$, because the
characteristic scale, $\bar k$, of $k$ in $r(k,t)$ is much larger than
$k_c$. 
Thus, the conservation identity \eqref{eq:conservation:equil} takes the
asymptotic form 
\begin{equation}
  \rho \approx \rho_c + \eta\rho_s\rho'(\rho_s)+\int_0^{\makebox[0mm][l]{$\scriptstyle\, \infty$}}
  k\,r(k,t)\,dk. \label{eq:continuum:conserv}
\end{equation}
Notice that if $r\equiv0$, which corresponds to negligible
super-critical clusters, \eqref{eq:continuum:conserv} reduces to \eqref{eq:eta:eff} with
$\eta=\eta_*$.
Hence, $\eta(t)$ has the effective initial condition $\eta(0) =\eta_*$.

In summary, the signaling problem for $r(k,t)$ consists of the advection PDE
\eqref{eq:continuum:PDE}, the creation BC \eqref{eq:continuum:BC}, and the functional dependence of
the super-saturation, $\eta$ on $r(k,t)$ in \eqref{eq:continuum:conserv}.
This signaling problem is nonlinear because $\eta$, as a functional of
$r$, appears in the advection PDE and, in addition, in the exponential
creation rate in the BC\@. 
\section{Conclusions}
The discrepancy between the two accepted model of nucleation---the surface reaction
model derived by Becker and D\"oring, and the diffusion limited growth
model due to Lifshitz and Slyozov---has now been resolved.
Our new model, in which  clusters interact with diffusing monomers 
by a surface reaction, predicts both models as limit cases. 
Although it provides a growth rate for clusters of all sizes, 
from small to large,  its limiting behaviors are of special interest for us. 
In the limit of small clusters, the BD model emerges, and we can derive
the Zeldovich creation rate of super-critical clusters using the BD kinetics.
In the limit of large clusters, the diffusion limited growth model
emerges, from which we derive the same evolution PDE for the density
function, $r(k,t)$, as LS.

Not all our findings corroborate the classical assumptions and results.
We find that the `common wisdom' about the  monomer
density at the cluster surface is wrong. 
In the classical DLG model, the surface density is chosen so that the
surface reaction is in equilibrium. 
In our model, the balance between the surface reaction and 
diffusion leads to a surface monomer density substantially different
from the `equilibrium' value.  
Despite this, the new model still predicts the same growth rate as the
classical DLG model (for large clusters).

Another piece of `common wisdom' is that BD pertains to small clusters and DLG to large ones. 
In our model, there is a new characteristic cluster size, $k_*$, that
separates small ($k\ll k_*$) from large ($k\gg k_*$). 
This is useful in providing definite predictions for the validity of
the model. 
For example, since nucleation happens around the critical size $k_c$, and the
Zeldovich creation rate assumes the small cluster limit of the model, the
validity of the Zeldovich formula requires $k_c\ll k_*$. 
The case where $k_c$ is of the same order or larger than $k_*$ is not
covered by the current paper nor by classical Zeldovich nucleation
theory. 
A separate investigation is required for this case, one that goes beyond  or treatment of the  monomer bath as a smooth `mean field' and really accounts for its discrete and fluctuation nature.

The monomer density around the cluster satisfies  the diffusion
equation and a mixed boundary condition at the surface of the
cluster. 
To calculate the flux of the monomers into or out of the cluster,
our model assumes that the monomer density in the vicinity of the
cluster is a quasi-static solution of the diffusion equation.
We show that the quasi-static `surface' value of monomer concentration
in our model is stable: starting from general initial conditions, the concentration of the monomer bath rapidly relaxes to the quasi-static approximation.

The preceding insights are codified in a \emph{signaling problem} for the cluster size distribution.
It consists of an advection PDE consistent with diffusion limited growth, and a BC consistent with the Zeldovich creation rate.
The PDE and BC contain the super-saturation as a parameter.
The super-saturation, in turn, is a function of the cluster size distribution as dictated be the overall conservation of particles.
Hence, the full signaling problem is non-linear. 
In particular, the Zeldovich creation rate is \emph{exponentially} small as the super-saturation goes to zero.
Therefore, accounting for small corrections to the monomer density is
important, especially in the initial stage when new clusters are
being created. 

Given an initial condition of pure monomers, quasi-static densities  are
formed during the ignition phase \cite{NBC05}, whereby small,
sub-critical clusters are created from the monomers. 
The ignition transient is a precursor to the nucleation of
super-critical clusters.
Therefore, once the nucleation `has ignited', the  monomers density is 
lower than the its original value, prior to the ignition transient.
The signaling problem starts from an effective IC that explicitly accounts for the loss of monomers during ignition.
This is especially relevant  in our case, where the
monomer-density is close to the saturation density, $\rho_s$, and the dependence of the PDE and BC on the monomer density is very strong. 

Such is the state of the \emph{theory}.
We now conclude the conclusion by a small trespass into the domain of
accountability: do the material parameters of real aggregation
processes cooperate with the various assumptions of the modeling?
We start with the characteristic cluster size $k_*$, most
conspicuously present in our model.
We examine physical constants associated with an aqueous solution of calcium
carbonate, CaCO\subs{3}.
Its solubility is rather small, so the basic requirement of
`diluteness' is satisfied.
Admittedly, CaCO\subs{3} dissolves into positive (Ca\sups{2+}) and
negative ($\text{CO}^{2-}_3$) ions, which is not reflected  in our
aggregation model with identical particles.
But we are examining \emph{crude} order of magnitude estimates, so
will not be distracted by the inconvenience, and we formally consider
a positive-negative ion pair as a `monomer.'

Formula \eqref{eq:k_*} for $k_*$ contains molecular volume $\nu$, the
saturation density of monomer $\rho_s$, the diffusion coefficient $D$ of
monomer in solution, and the dissociation rate constant $\omega$.
Estimates of $\nu, \rho_c,$ and $D$ are readily found in a
chemical handbook \cite{CRCTable}.
We use $\rho_c$ as an order of magnitude estimate of $\rho_s$.
The dissociation rate, $\omega$, is much more elusive.
Here, we indulge in the activation energy model similar to Kelton's,
in his review of glass-crystal transitions \cite{KELT91}.
The model is summarized by the formula
\begin{equation}
  \omega = \frac{D}{\nu^{2/3}}e^{-\beta},\label{eq:kelton:formula}
\end{equation}
Here, $\frac{\nu^{2/3}}{D}$is the characteristic time for a monomer to
diffuse a distance comparable to its own size, and $\beta$ is an
`activation energy of dissociation', in units of $k_BT$.
Inserting \eqref{eq:kelton:formula} for $\omega$ into \eqref{eq:k_*},
we find
\begin{align}
  k_*&=(3\cdot 16\pi^2) \brk{e^\beta\rho_s}^3,\notag
\intertext{or, using $\rho_s=e^{-\alpha}$,}
 k_*&=(3\cdot 16\pi^2) e^{3\brk{\beta-\alpha}}.\label{eq:k_*:2}
\end{align}

Formula \eqref{eq:kelton:formula} has other  applications for us,
besides the estimate \eqref{eq:k_*:2} of $k_*$.
For instance, recall that the quasi-static limit of the FBP \eqref{eq:eta:lap} is based on the `diffusion time' $\frac{\nu^{2/3}}{D}$ much smaller than the dissociation time $\frac{1}{\omega}$, so we required $\frac{D}{\omega\nu^{2/3}}\gg1$.
From \eqref{eq:kelton:formula} we get 
\begin{equation}
  \frac{D}{\omega\nu^{2/3}} = e^\beta\label{eq:beta:ratio}
\end{equation}
and so an activation energy $k_BT\beta$ on the order of a few $K_BT$ is sufficient.

Next, we examine the criterion $\frac{k_c}{k_*}\ll1$ for the validity of the Zeldovich nucleation rate.
Using (\ref{eq:critical:k:2}, \ref{eq:k_*:2}), we find this implies a bound on the super-saturation,
\begin{equation*}
  \eta\gg \sig \,e^{\alpha-\beta}.
\end{equation*}
Using the crude `cube' model of bonding energy \eqref{eq:energy:cube} we estimate $\sig$ by, $\sig\approx \frac23\alpha$ so the criterion on 
the supersaturation becomes
\begin{equation}
  \eta \gg \alpha \,e^{\alpha-\beta}.\label{eq:eta:lower:bound}
\end{equation}

A lower bound on the super-saturation might seem like a problem, as we expect $\eta$ to asymptote to zero in the long-time limit of an aggregation process.
However, significant nucleation occurs at an early and relatively brief phase, so \eqref{eq:eta:lower:bound} should apply for the initial super-saturation. 
In late stage coarsening, $\eta$ is  much smaller than the RHS of \eqref{eq:eta:lower:bound}, but significant nucleation is not happening then.

Numerical evaluation of $k_*$, or the RHS of \eqref{eq:eta:lower:bound} require actual values of $\alpha$ and $\beta$.
It is relatively easy to find $\alpha$, as we have $\alpha = -\log \rho_s$ and $\rho_s$ is essentially the volume fraction of monomer in saturated solution.
For instance, from the data in \cite{CRCTable}, we find that $2.44\times10^{-4}\text{cm}^3$ of solid CaCO\subs{3} is soluble in 100cm\sups{3} of water at room temperature. 
The volume of the solution is nearly 100cm\sups{3} so the volume fraction occupied by Ca\sups{2+}, $\text{CO}^{2-}_3$ is (roughly) $\rho_s=2.44\times10^{-6}$, and our estimate of $\alpha$ is $\alpha\approx12.9$.
The bad news is that we do not know $\beta$ any better than we know $\omega$ in the first place, so we cannot do primia-facie evaluations of $k_*$ nor the RHS of \eqref{eq:eta:lower:bound}.

Our policy is to use  \eqref{eq:eta:lower:bound} to obtain bounds on $\beta$ for which our model is valid.
For instance, we have seen that the quasi-static approximation requires $\beta\gg1$.  
In addition, our whole analysis is based on small super-saturation, so $\eta\ll1$.
This is compatible with \eqref{eq:eta:lower:bound} only if 
\begin{equation}
  \alpha e^{\alpha-\beta}\ll1.\label{eq:beta:bound:2}
\end{equation}
Starting with $\alpha=12.9$, we find the LHS is unity if
$\beta-\alpha=2.56$, but if we increase the activation energy, $\beta$, by $3\,k_BT$ we get $\beta-\alpha=5.56$ and  the LHS of \eqref{eq:beta:bound:2} is $0.061\ll1$.
Inserting the value of $\beta-\alpha=5.56$ into \eqref{eq:k_*:2} for $k_*$ results in $k_*\approx8.3\times10^9$.

Our estimate of $k_*$ does not really use values of $D$ and $v$ because the combination $\frac{D^3}{v^2}$, which appears in the formula \eqref{eq:k_*} is absorbed by the Kelton formula \eqref{eq:kelton:formula} for $w$.
Again we `turn the equations around': using physical values of $D, v$, and $\rho_s$, and our `made up' estimate of $k_*$, we work backwards to $w$.
From the Handbook of Chemistry and Physics \cite{CRCTable}, we find that the diffusion coefficients of Ca\sups{2+} and $\text{CO}^{2-}_3$ are $.792\times10^{-5}\text{cm}^2/\text{s}$ and $.9.23\times10^{-5}\text{cm}^2/\text{s}$. 
So, as a crude estimate we put $D\approx 10^{-5} \text{cm}^2/\text{s}$.
The nucleous colume of solid CaCO\subs{3} is $v = 4.51\times10^{-26}\text{cm}^3$.
Inserting these values of $D$ and $v$ into \eqref{eq:k_*}, along with the previously determined $\rho_s$ and $k_*$ we obtain the estimate of $w$, $w\approx2\times10^6\text{s}^{-1}$.

\section{Acknowledgments}
This paper was supported in part by the National Science Foundation.
Y. Farjoun was partially supported by grant number DMS--0515616.
\appendix
\section{Appendix}

\subsection{Deterministic Cluster Growth According to BD}
\label{sub:sec:det:BD}
Here we determine the domain of $k$ in which the average change
in the cluster size $\Average{\delta k}$ is a good approximation to the actual
change in cluster size. 
BD assumes that the cluster grows and shrinks by means of discrete,
independent Poisson processes. 
The cluster gains monomers at a rate $c_k\rho_1$ via the adsorption process, and via the emission process it loses monomers with rate
$d_k$. 
Therefore, the average change in size over a short interval $\delta t$ is 
\begin{equation*}
  \Average{\delta k} = \Average{k(t+\delta t)-k(t)} = (c_k\rho_1-d_k)\delta t,
\end{equation*}
provided that $\delta t$ is small enough so that 
\begin{equation}
\Abs{c_k\rho_1-d_k}\delta t \ll k.\label{eq:delta:k}
\end{equation}

Since the two controlling process are assumed to be independent
Poisson processes, the mean square deviation of $k$ from the average
is 
\begin{equation*}
  \Average{(\delta k -\Average{\delta k})^2} =  (c_k\rho_1+d_k)\delta t
\end{equation*}
The process can be considered deterministic if the deviation is much
smaller than the expected change, so
\begin{equation}
    \sqrt{c_k\rho_1+d_k}\sqrt{\delta t}\ll \Abs{(c_k\rho_1-d_k)\delta t}.\label{eq:deterministic}
\end{equation}
Combining inequalities \eqref{eq:delta:k} and \eqref{eq:deterministic}
we get a condition on $\delta t$
\begin{equation*}
    c_k\rho_1+d_k \ll \Abs{(c_k\rho_1-d_k)}^2 \delta
    t\ll\Abs{c_k\rho_1-d_k}k.
\end{equation*}
We can find $\delta t$ which satisfies this condition if 
\begin{equation}
    c_k\rho_1+d_k \ll\Abs{c_k\rho_1-d_k}k.\label{eq:deterministic:con}
\end{equation}
This does not hold for every value of $k$. 
In particular, we notice
that for $k=k_c$ the RHS is zero. 
Hence, the $k(t)$ that can be approximated as deterministic must be
bounded away from $k_c$. 

For $k\gg1$ in \eqref{eq:deterministic:con} we insert the approximations (\ref{eq:d_k}, \ref{eq:detailed:balance}) for $d_k$ and $c_k$, and then examine its asymptotic form as $\eta\goto0^+$. 
The result is 
\begin{equation}
  \brk{2+\eta\brk{1-\brk{\frac{k}{k_c}}^{-\third}}}\ll \eta\Abs{1-\brk{\frac{k}{k_c}}^{-\third}}k.\label{eq:det:con:2}
\end{equation}

For clusters with $k\gg k_c$, this condition is satisfied trivially for $\eta\ll1$. 
For $k$ on the order of $k_c$, we need to dig a little deeper.

Since we are looking at the $\eta\goto0$ limit, we look at the leading order terms of \eqref{eq:det:con:2}
 \begin{equation*}
  \frac2{\eta k}\ll \Abs{1-\brk{\frac{k_c}{k}}^{\third}}.
\end{equation*}
For $k$ near $k_c$ we expand the RHS around $k=k_c$.
Keeping the leading order terms of the Taylor series of the RHS, we
find an explicit condition for the evolution of clusters to be treated
deterministically: $k$ must be bounded away from $k_c$ according to
\begin{equation*}
  \frac6\eta\ll\Abs{k-k_c}.
\end{equation*}
As $\eta\goto0$, the excluded domain grows.
On the other hand, its  size relative to $k_c\sim\brk{\frac{\sig}{\eta}}^3$ shrinks to zero: 
\begin{equation*}
  \frac{6\eta^2}{\sig^3}\ll\Abs{\frac{k-k_c}{k_c}}.
\end{equation*}
Therefore, we see that for small super-saturation the evolution of
clusters can be treated deterministically, for all but a vanishingly small domain around $k_c$. 
\subsection{Effective Super-Saturation, $\eta$, After Ignition}
\label{sub:sec:eta:eff}
It has previously been shown that an initial condition of pure monomer
leads to a transient `ignition' phase in which quasi-static
densities of sub-critical clusters are created \cite{NBC05}.
Since these sub-critical clusters are made of monomers, after the
ignition transient the monomer density will be lower than the
original value, prior to the ignition. 
We compute the value $\eta_*$ of super-saturation after the formation of the $k<k_c$ quasi-static densities, but before significant depletion  by super-critical clusters.

The starting point if the sum \eqref{eq:conservation} truncated to $k\le k_c$ because the contribution from super-critical clusters is insignificant:
\begin{equation*}
  \rho=\sum_{\makebox[0mm]{$\scriptstyle 1\le k\le k_c$}} k\rho_k.
\end{equation*}
Since the sub-critical cluster densities are quasi-static, we
approximate them by the equilibrium densities \eqref{eq:rho_k:equil}, and the corresponding approximation of the sum is
\begin{equation*}
  \rho=\sum_{\makebox[0mm]{$\scriptstyle 1\le k\le k_c$}} k\rho_1^ke^{-\frac{\eps_k}{k_BT}}.
\end{equation*}
Recalling the definition of the super-saturation,
\eqref{eq:supersaturation},  $\rho_1 = \rho_s (1+\eta_*)$, and taking
the two-term expansion as $\eta_*\goto0^+$, we find
\begin{equation*}
\rho\sim\rho_c+\eta_*\sum_{\makebox[0mm]{$\scriptstyle 1\le k\le k_c$}}k^2\rho_s^ke^{-\frac{\eps_k}{k_BT}}.
\end{equation*}
The sum on the RHS can be obtained by differentiating the series
representation  \eqref{eq:conservation:equil} of $\rho(\rho_1)$, setting $\rho_1=\rho_s$, multiplying by $\rho_s$, and then truncating to $k\le k_c$.
The derivative series converges at $\rho_1=\rho_s$, so for $\eta\goto0$ (and hence for $k_c\goto\infty$), the sum is asymptotic to $\rho_s\rho'(\rho_s)$.
Hence,
\begin{equation*}
  \rho\sim\rho_c+\eta_*\rho_s\rho'(\rho_s), 
\end{equation*}
and so,
\begin{equation*}
  \eta_* = \oneover{\rho'(\rho_s)} \frac{\rho-\rho_c}{\rho_s}.
\end{equation*}

\bibliography{general}

\end{document}